\documentclass[fleqn,12pt,twoside]{article}
\usepackage{espcrc1}
\usepackage{graphicx}

\newcommand{\AmS}{{\protect\the\textfont2
  A\kern-.1667em\lower.5ex\hbox{M}\kern-.125emS}}
\title{Stability and structure of quark matter
       in a molecular dynamics framework}
\author{Yuka Akimura\address[saitama]{Department of physics,
        Saitama University, 
        Sakura-Ku, Saitama-Shi, 338-8570, Japan
        }\address[JAERI]{Advanced Science Research Center,
        Japan Atomic Energy Research Institute, Tokai, Ibaraki 319-1195,
        Japan}\ ,
        Toshiki Maruyama\addressmark[JAERI]\ ,
        Naotaka Yoshinaga\addressmark[saitama]\ , 
        and  Satoshi Chiba\addressmark[JAERI] }
\begin{document}
\maketitle

\begin{abstract}
We study stability and structure of quark matters 
as a function of density
in a framework of molecular dynamics (MD).
Using appropriate effective interactions and the frictional 
cooling method, we search for the minimum energy of the system.
Transition from confined to deconfined phase is observed at 
densities of 2 -- 3$\rho_0$, 
where $\rho_0$ is the nuclear matter saturation density.
The $uds$ matter becomes more stable than the 
charge-neutral $ud$ matter at 3$\rho_0$,
but the $udd$ matter is the most stable even at high density.
\end{abstract}

\section{Introduction}
More than twenty years ago Bodmer and later Witten put a hypothesis that
the absolute ground state of strongly interacting matter is the deconfined
state of quark matter consisting of an equal proportion of up, down,
and strange quarks called strange quark matter even at zero temperature 
and zero pressure \cite{Bodm71,witten}.

Recently some exotic baryon systems 
including strange quarks were discovered.
For example, the penta-quark, called ${\it \Theta^+}$,
consisting of 5 quarks ($\it{uudd\bar{s}}$)
with an extraordinary small width $\sim$25MeV, 
was discovered at SPring8 \cite{Nakn03}.
At KEK, a strongly bounded kaonic system, 
$ppnK^-$ \cite{Iwas03,suzuki}, which was predicted
from an AMD calculation \cite{Dote03} 
to form a high density system of 10$\rho_0$, was discovered.
 From an astrophysical point of view,
there is a possibility that the strange quark matter is 
realized at the center of neutron stars and/or quark stars \cite{Drak02}.
At high baryon density such as in the core of these stars,
quark gluon plasma (QGP) may exist and
confined baryon and deconfined baryon states may coexist \cite{Mish00,Sanj98}.
Searching for the energy region in which the hadron 
to quark transition occurs and the strange quark matter becomes stable 
is one of the most interesting topics in high energy hadron physics.

In this paper we study the stability of quark matter 
at finite baryon density at zero temperature
using a molecular dynamics method (MD).
In this framework it is not necessary to assume 
whether the state of quarks is confined or
deconfined and their mixed state can also be treated \cite{Maru00}.

\section{Molecular dynamics for quark system}

We define the total wave function of a system $\Psi$ with baryon number $A$
as a direct product of single particle wave functions 
with a Gaussian wave packet in coordinate space
and a state vector $\chi_i$ with a fixed spin orientation, flavor and color,
\begin{equation}
\Psi=\prod_{i=1}^{3A} \frac{1}{(\pi L^2)^{3/4}}
\exp \left[ -\frac{({\bf r}-{\bf R_{\it i}})^2}{2L^2} 
+\frac{i}{\hbar}{\bf P}_i{\bf r}
\right]\chi_i.
\end{equation}
The antisymmetrization of the wave function is neglected, but treated 
phenomenologically by introducing a Pauli potential
which acts as a repulsive force between quarks of the same
color, flavor and spin orientation.
%
\begin{figure}
\includegraphics[width=0.85\textwidth]
{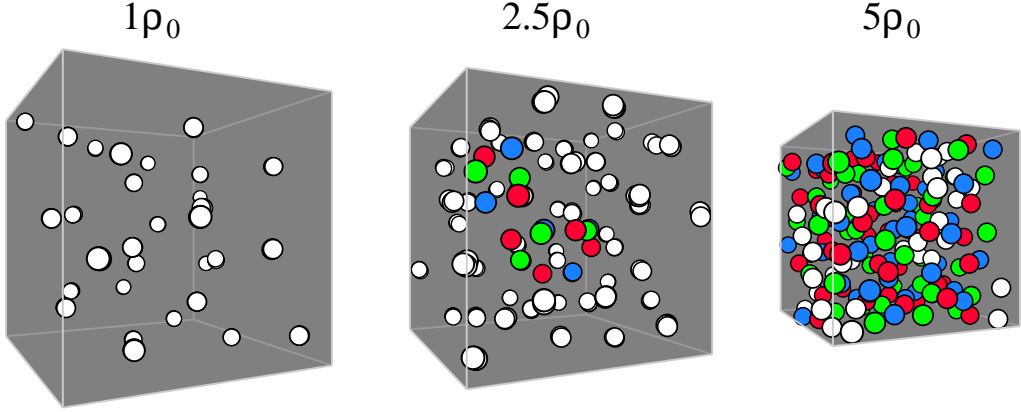}
\caption{Snapshots for the $udd$ matter after cooling.
The white circles denote confined quarks and colored circles
are deconfined quarks.}
\end{figure}
%
The Hamiltonian is expressed as $H=H_0+V_{\rm Pauli}-T_{\rm spur}$,  
where $T_{\rm spur}$ is a spurious zero point kinetic energy of clusters 
as described in \cite{Ono92}. 
The original Hamiltonian $H_0$ and $V_{\rm Pauli}$ 
are explicitly expressed as,
\begin{eqnarray}
H_0&=&
\left<\Psi\biggm|
\sum_{i} \left( \frac{{\hat{\bf p}}_{i}^2}{2m_i} + m_i \right) +
\frac{1}{2}\sum_{j \neq i}\left(-\sum_{a=1}^8 
\frac{\lambda_i^a \lambda_j^a}{4}
\left(K\hat{r}-\frac{\alpha_s}{\hat{r}}
\right)
+\hat{V}_{\rm meson}
\right)
\biggm|\Psi\right>
,\\
\hat{V}_{\rm meson}&\equiv&-\frac{g_{\sigma q}^2}{4 \pi }
\frac{e^{- \mu _{ \sigma } \hat{r}}}{\hat{r}}
+\frac{g_{\omega q}^2}{4 \pi }
\frac{e^ {- \mu _{ \omega } \hat{r}}}{\hat{r}}, \\
V_{\rm Pauli}&\equiv&
\frac{1}{2}\sum_{j \neq i} 
 C_{p} \left ( \frac{\hbar}{q_0 p_0} \right )^3
\exp \left[-\frac{({\bf R_{\it i}}-{\bf R_{\it j}})^2}{2q_0^2}
-\frac{({\bf P_{\it i}}-{\bf P_{\it j}})^2}{2p_0^2}
\right]  , 
\end{eqnarray}
where $\hat{r}=|\hat {\bf r}_i-\hat {\bf r}_j|$
and $\lambda^a_i=\left<\chi_i|\lambda^a|\chi_i\right>$. 
We use $\alpha_s$=1.25 (the QCD fine structure constant),
$K$=900MeV/fm (string tension), constituent quark masses
$m_{u,d}$=300MeV and $m_s$=500MeV throughout this simulation
which are typical values of parameters in quark models. 
The product of Gell-Mann matrices for different colors 
and for same colors are
$\left<\sum_{a=1}^8 \lambda_i^a \lambda_j^a\right>= -\frac23$
and $\frac43$, respectively.
However, since the antisymmetrization is neglected in our model, 
we multiply those values by 4 to effectively include 
the exchange term of color interaction. 
To introduce the nuclear force between white (colorless) baryons,
we employ the meson exchange potential ${\hat V}_{\rm meson}$ 
acting between quarks.
The $\sigma$ and $\omega$ meson-quark coupling constants 
are $g_{\sigma q}$=3.42 and $g_{\omega q}$=9.0, 
the meson masses are $\mu_\sigma$=300MeV
and $\mu_\omega$=782MeV, respectively.
The parameters in the Pauli potential $V_{\rm Pauli}$ are 
determined to reproduce the kinetic energy of 
the Fermi gas \cite{Maru97}.
In the present study we adopt $C_p$=131MeV,
$q_0$=1.8fm, $p_0$=120MeV for {\it u} and {\it d} quarks and
$C_p$=79MeV for {\it s} quarks.
The size of the quark wave packet, $L$, is taken to be 0.43fm,
whereas $L^{\rm eff}$ =0.7fm is used in $\left<{\hat V}_{\rm meson}\right>$
to take into account of the baryon size \cite{Maru00}.

\section{Quark matter properties}

Here we investigate the ground state (zero temperature) of matter
consisting of quarks.
As an initial condition we distribute ``hadrons'' 
(made from 3 quarks with different colors) 
randomly in a cell with a periodic boundary condition.
In the case of baryon density, e.g., of 1$\rho_0$, 
90 quarks are put in a cell of a size 5.6 fm.
To search for the minimum energy configuration of the system,
we solve the cooling equations of motion 
with friction terms as in \cite{Maru97}.
The simulations are performed for ``$ud$ matter'' 
(with same numbers of $u$ and $d$ quarks), 
``$udd$ matter'' (number of $d$ quarks is twice that of $u$ quarks)
and ``$uds$ matter'' (with same numbers of $u$, $d$, and $s$ quarks).

\begin{figure}
\begin{minipage}[c]{0.58\textwidth}
\includegraphics[width=\textwidth]
{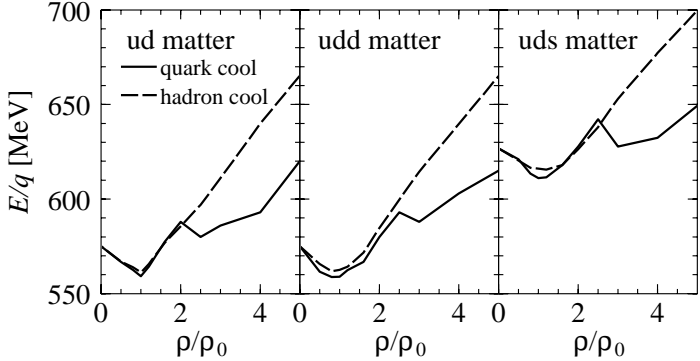}
\end{minipage}
\hspace{\fill}
\begin{minipage}[c]{0.36\textwidth}
\vspace{-1cm}
\caption{Energy per quark for three kinds of matter.
Dashed lines denote the result of ``hadron-cooling''
in which 3 quarks always form a colorless hadron,
 while solid lines are 
those of ordinary ``quark-cooling'' without any constraint.}
\end{minipage}
\end{figure}

Snapshots of the $udd$ matter for different baryon densities 
are displayed in Fig.~1. 
All quarks are confined as hadrons at baryon density of 1$\rho_0$.
At $2.5\rho_0$, a partly deconfined state can be seen, 
whereas almost all quarks are deconfined at $5\rho_0$.

The ground state energies of matter are shown with
solid lines in Fig.~2.
The dashed lines show results with a constraint that 
${\bf R}_i$ of 3 quarks in each initial ``hadron''
should always be equal (to keep hadrons by hand). 
Here we call this calculation ``hadron-cooling'',
while the normal one we call ``quark-cooling''.
At lower density, the energy of the system by hadron-cooling 
agrees that by quark-cooling.
This means that quarks are confined in hadrons as seen in Fig.~1.
Note that the saturation of hadronic matter is 
realized around 1$\rho_0$ for the first time
in this kind of quark MD simulation.
The energies of the $ud$ matter and the $udd$ matter are almost the same.
This is because the Pauli potential is not fully effective for ``hadrons''
and $\rho$ meson exchange is missing.
At densities of 2 -- 3$\rho_0$, transition to deconfined phase occurs.
In our simulation, the decrease of the kinetic energy and the increase 
of the potential energy act oppositely during the deconfinement.
The softening of the latter effect may be the origin of the transition
at higher densities.

Figure 3 shows comparison of energy among 
the $ud$ matter with relativistic electrons, 
the $udd$ matter and the $uds$ matter under charge neutral condition.
Although the $uds$ matter is the most unstable at low density,
it becomes more stable than the $ud$ matter 
as deconfinement occurs at 3$\rho_0$. 
The $udd$ matter, however, is the most stable 
over all ranges in our present result.
%
\begin{figure}
\begin{minipage}[c]{0.47\textwidth}
\includegraphics[width=\textwidth]
{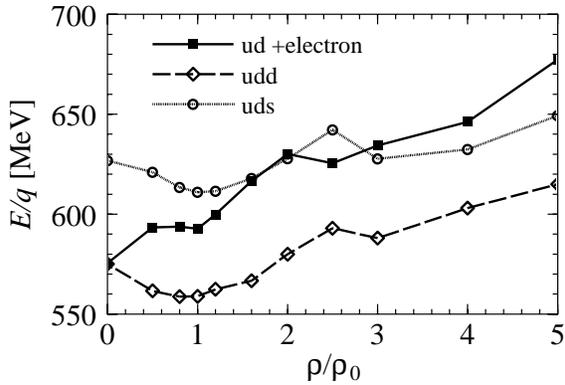}
\end{minipage}
\hspace{\fill}
\begin{minipage}[b]{0.43\textwidth}
\vspace{-4cm}
\caption{Baryon density dependence of the energy per quark
for three kinds of matter under charge-neutral condition.}
\end{minipage}
\end{figure}

\section{Summary}

Quark many-body systems were studied in a framework of MD.
It has been found that hadron matter undergoes a phase
transition to the quark matter as baryon density increases.
At low density, the saturation for the $ud$ matter is realized
by introducing ${\hat V}_{\rm meson}$.
The $udd$ and $uds$ matter, however, also show the similar 
saturation because exactly the same meson exchange potentials 
are used among all quarks.
To improve this, a flavor-dependent potential
like $\rho$ meson exchange potential is needed.
The $uds$ matter becomes more stable than 
the charge-neutral $ud$ matter at 3$\rho_0$.
But the $udd$ matter is the most stable even at high density.
We expect more reasonable results if $\rho$ meson exchange 
potential is included in the future.


\end{document}